\documentstyle[a4,12pt,amsmath,amssymb,latexsym,amssymb,natbib]{article}
\def \dd{{\mathrm{d}}}

\def \pd{\partial}

\def \tl#1{\overset{\kern 1pt\circ}{#1}}
\def \TL#1{\overset{\kern -3pt \circ}{#1}}
\def \TLL#1{\overset{\kern -7pt \circ}{#1}}

\begin{document}
\title{{\bf The Eshelby stress tensor, angular momentum tensor and scaling
    flux in micropolar elasticity}}
\author{Markus Lazar~$^\text{a,}$\footnote{
Corresponding author. {\it E-mail address:} lazar@fkp.tu-darmstadt.de (M.~Lazar).}\; ,
Helmut O.K. Kirchner~$^\text{b,c,}$\footnote{{\it E-mail address:} kirchnerhok@hotmail.com (H.O.K.~Kirchner).}
\\ \\
${}^\text{a}$ 
        Emmy Noether Research Group,\\
        Department of Physics,\\
        Darmstadt University of Technology,\\
        Hochschulstr. 6,\\      
        D-64289 Darmstadt, Germany\\
${}^\text{b}$ 
        Universit{\'e} Paris-Sud, UMR8 182, Orsay, F-91405\\     
${}^\text{c}$ 
        CNRS, Orsay, F-91405
}

\date{\today}    
\maketitle

\begin{abstract}
The (static) energy momentum tensor, angular momentum tensor
and scaling flux vector of micropolar elasticity are derived within the framework
of Noether's theorem on variational principles. 
Certain balance (or broken conservation) laws of broken translational, rotational
and dilatational symmetries are found including inhomogeneities,
elastic anisotropy, body forces, body couples and dislocations and disclinations
present. 
The non-conserved $J$-, $L$- and $M$-integrals of micropolar elasticity 
are derived and discussed. 
We give explicit formulae for the configurational forces,
moments and work terms.
\\

\end{abstract}
\hspace{6mm}

\section{Introduction}
Symmetries and conservation laws of micropolar elasticity are 
of important interest in mathematical physics, material science 
and engineering science.
\citet{Jaric,Jaric2} and \citet{Dai} studied conservation laws in micropolar elastostatics.
\citet{Vuko} obtained some conservation laws for micropolar elastodynamics.
The Noether theorem was applied by~\citet{PS,NZ,Maugin98,LM03} to obtain 
conservation laws and the corresponding conserved currents for linear
micropolar elasticity. For couple stress elasticity the Noether theorem was used
by~\citet{LM00}. All these investigations were restricted to homogeneous, source-free and compatible
micropolar elasticity. 
The derived conservation laws correspond to variational invariance 
of the strain energy with respect to translation and rotation symmetries.
Therefore, the $J$- and $L$-integrals are conserved in 
such a version of micropolar elasticity.
On the other hand, the scaling symmetry is not a variational symmetry in
micropolar elasticity, thus, the $M$-integral is not conserved (see, e.g., \citet{LM03}).

Not so many results on conservation laws are known for micropolar 
elasticity with defects.
Only the Eshelby stress tensor and the configurational forces caused by defects 
are known~\citep{Kluge,Kluge2}.
The Eshelby stress tensor corresponds to translation symmetry and 
it may be identified with the (static) 
energy-momentum tensor (EMT).
But nothing is known in the literature for nonhomogeneous
micropolar elasticity with body forces, body couples, dislocations and
disclinations present.

It is the purpose of the present paper to extend the results of these earlier
results on micropolar elasticity to account for material nonhomogeneity, anisotropy, defects, body
forces and body couples.  
We derive balance laws breaking the translation, rotation and scaling symmetries.
The symmetry breaking terms are called configurational forces, configurational
moments and configurational work. 
In turn, we find the expressions for the Eshelby stress tensor, 
angular momentum tensor and scaling flux in micropolar elasticity.

\section{Basic equations of micropolar elasticity}
\setcounter{equation}{0}
\label{elasticity}
In this section, 
we recall the basics of micropolar elasticity~\citep{Eringen99}.
We consider the general case of 
anisotropic linear micropolar elasticity theory 
for non-homogeneous and incompatible media with defects.
The strain energy for a  micropolar material reads
\begin{align}
\label{en-grad1}
W=\int\, w\, \dd V
\end{align}
with the energy density
\begin{align}
\label{en-grad2}
w=\frac{1}{2}\, A_{ijkl}\gamma_{ij} \gamma_{kl}
+B_{ijkl} \gamma_{ij}\kappa_{kl}
+\frac{1}{2}\, C_{ijkl} \kappa_{ij}\kappa_{kl} ,
\end{align}
where $\gamma_{ij}$ denotes the elastic micropolar distortion 
tensor and $\kappa_{ij}$ is the elastic wryness tensor.
These elastic `strain' tensors are given in terms of a displacement vector
$u_i$ and a microrotation $\phi_i$. Additionally, the total `strains' may be decomposed
into elastic and plastic parts according to
\begin{align}
\label{deco1}
\gamma_{ij}^{\text{T}}&=\pd_j u_i +\epsilon_{ijk}\phi_k
=\gamma_{ij}+\gamma^{\text{P}}_{ij},\\
\label{deco2}
\kappa^{\text{T}}_{ij}&=\pd_j \phi_i
=\kappa_{ij}+\kappa^{\text{P}}_{ij}.
\end{align}
Here $\gamma^{\text{P}}_{ij}$  is the plastic distortion and $\kappa^{\text{P}}_{ij}$
is the plastic wryness.
For simplicity, we have assumed a linear relationship but that is not at all necessary.
The constitutive relations for full anisotropy read:
\begin{align}
\label{CR1}
t_{ij}&=\frac{\pd w}{\pd \gamma_{ij}}=A_{ijkl} \gamma_{kl}+B_{ijkl}\kappa_{kl} ,\\
\label{CR2}
m_{ij}&=\frac{\pd w}{\pd \kappa_{ij}}=B_{klij}\gamma_{kl}+C_{ijkl}\kappa_{kl} ,
\end{align}
where $A_{ijkl}$, $B_{ijkl}$ and $C_{ijkl}$ are the elastic tensors of micropolar elasticity
with the symmetries
\begin{align}
\label{symm}
A_{ijkl}=A_{klij},\qquad
C_{ijkl}=C_{klij}.
\end{align}
Dimensionally, $[C_{ijkl}]=\ell\, [B_{ijkl}]=\ell^2\, [A_{ijkl}]$, where $\ell$
is a material length parameter.
For the non-homogeneous medium under consideration, they depend on position,
$A_{ijkl}(x)$, $B_{ijkl}(x)$ and $C_{ijkl}(x)$.
$t_{ij}$ is the force stress tensor and $m_{ij}$ is the couple stress tensor.
For an isotropic micropolar material the elastic tensors
simplify to
\begin{align}
\label{CR-Iso}
A_{ijkl}&=\lambda\, \delta_{ij}\delta_{kl}
+\mu\big(\delta_{ik}\delta_{jl}+\delta_{il}\delta_{jk}\big)
+\mu_c\big(\delta_{ik}\delta_{jl}-\delta_{il}\delta_{jk}\big),\nonumber \\
C_{ijkl}&=\alpha\, \delta_{ij}\delta_{kl}
+\beta\big(\delta_{ik}\delta_{jl}+\delta_{il}\delta_{jk}\big)
+\gamma\big(\delta_{ik}\delta_{jl}-\delta_{il}\delta_{jk}\big), \\
B_{ijkl}&=0\, ,\nonumber
\end{align}
where $\mu$ is the shear modulus, $\lambda$ denotes the Lam{\'e} constant,
and $\mu_c$, $\alpha$, $\beta$ and $\gamma$ are additional material constants 
for micropolar elasticity. 

The field equations in presence of an external force $f_i$ and an external couple
$l_i$ are given by
\begin{align}
\label{fe1}
\pd_j t_{ij}+f_i&=0,\\
\label{fe2}
\pd_j m_{ij}-\epsilon_{ijk} t_{jk}+l_i&=0.
\end{align}
In linear micropolar elasticity, the incompatibility equations are
\begin{align}
\label{Disl0}
\epsilon_{jkl}(\pd_k\gamma_{il}+\epsilon_{ikm}\kappa_{ml})&=\alpha_{ij},\\
\label{Disc0}
\epsilon_{jkl}\pd_k\kappa_{il}&=\Theta_{ij},
\end{align}
and 
\begin{align}
\label{Disl2}
-\epsilon_{jkl}(\pd_k\gamma^{\text{P}}_{il}+\epsilon_{ikm}\kappa^{\text{P}}_{ml})&=\alpha_{ij},\\
\label{Disc2}
-\epsilon_{jkl}\pd_k\kappa^{\text{P}}_{il}&=\Theta_{ij},
\end{align}
where $\alpha_{ij}$ and $\Theta_{ij}$ are the dislocation density 
and disclination density tensors, respectively.
By differentiating we obtain the conservation laws for the dislocation density
and disclination density tensors:
\begin{align}
\pd_j\alpha_{ij}-\epsilon_{ijk}\Theta_{jk}&=0 ,\\
\pd_j \Theta_{ij}&=0,
\end{align}
which mean that the disclination density tensor is divergence free in the 
second index 
and the divergence of the dislocation density tensor 
is determined by the skew-symmetric part 
of the disclination density tensor.
Equivalently, upon multiplication of (\ref{Disl0}) and (\ref{Disc0}) with
the Levi-Civita tensor
\begin{align}
\label{Disl1}
\pd_k\gamma_{ij}-\pd_j\gamma_{ik}+\epsilon_{ikl}\kappa_{lj}-\epsilon_{ijl}\kappa_{lk}&=\epsilon_{kjl}\alpha_{il},\\
\label{Disc1}
\pd_k\kappa_{ij}-\pd_j\kappa_{ik}&=\epsilon_{kjl}\Theta_{il}.
\end{align}

\section{The Eshelby stress tensor and configurational forces 
in micropolar elasticity}
\setcounter{equation}{0}
\label{EMT-sec}

Let us take an arbitrary infinitesimal functional derivative $\delta W$
of the elastic energy density. 
We follow the procedure given by~\citet{Kirchner99} in order to construct
the energy-momentum tensor and the corresponding configurational forces.
From Eqs.~(\ref{en-grad1}), (\ref{en-grad2}) 
and (\ref{symm}) we get
\begin{align}
\label{W-var1-elast}
\delta W&=\frac{1}{2}\int\Big\{ [\delta A_{ijkl}]\gamma_{ij} \gamma_{kl}
+2 A_{ijkl} \gamma_{ij} [\delta \gamma_{kl}]
+2[\delta B_{ijkl}]\gamma_{ij} \kappa_{kl}
+2B_{ijkl} [\delta\gamma_{ij}]  \kappa_{kl}
\nonumber\\
&\qquad
+2B_{ijkl} \gamma_{ij} [\delta  \kappa_{kl}]
+ [\delta C_{ijkl}]\kappa_{ij}\kappa_{kl}
+2 C_{ijkl} \kappa_{ij} [\delta  \kappa_{kl}]
\Big\}\, \dd V  .
\end{align}
With the constitutive relations~(\ref{CR1}) and (\ref{CR2}) there remains
\begin{align}
\label{W-var}
\delta W&=\int\Big\{ 
t_{ij} [\delta \gamma_{ij}]+m_{ij} [\delta \kappa_{ij}] 
+\frac{1}{2}[\delta A_{ijkl}]\gamma_{ij} \gamma_{kl}
+[\delta B_{ijkl}]\gamma_{ij} \kappa_{kl}
+\frac{1}{2} [\delta C_{ijkl}]\kappa_{ij}\kappa_{kl}
\Big\}\, \dd V  .
\end{align}

Having configurational forces in mind, we specify the functional derivative 
to be translational:
\begin{align}
\label{transl}
\delta=(\delta x_k)\pd_k ,
\end{align} 
where $(\delta x_k)$ is an infinitesimal shift in the $x_k$-direction.
On the left hand side of Eq.~(\ref{W-var1-elast}) we write 
\begin{align}
\label{W-var-l-elast}
\delta W=\int \delta w\, \dd V=\int[\pd_k w](\delta x_k)\, \dd V
=\int\pd_i[w \delta_{ik}](\delta x_k)\, \dd V
\end{align}     
with the energy density~(\ref{en-grad2}). 
On the right hand side of Eq.~(\ref{W-var1-elast}) we have 
\begin{align}
\label{W-var-r-elast}
\delta W&=\int\Big\{
t_{ij} [\pd_k \gamma_{ij}-\pd_j\gamma_{ik}]+t_{ij}\pd_j\gamma_{ik}
+m_{ij} [\pd_k \kappa_{ij}-\pd_j\kappa_{ik}]+m_{ij} \pd_j\kappa_{ik}
\\
&\hspace{1.5cm}
+\frac{1}{2}\, \gamma_{ij}[\pd_k A_{ijmn}]\gamma_{mn}
+\gamma_{ij}[\pd_k B_{ijmn}]\kappa_{mn}
+\frac{1}{2}\, \kappa_{ij}[\pd_k C_{ijmn}]\kappa_{mn}
\Big\}
(\delta x_k)\, \dd V ,\nonumber
\end{align}
where the second, third, fifth and sixth terms have been 
subtracted and added.
The purpose is to obtain the square brackets with the meaning of Eqs.~(\ref{Disl1})
and (\ref{Disc1}). 
The third and sixth terms may be written 
\begin{align}
\label{W-v1}
t_{ij}[\pd_j \gamma_{ik}]&=\pd_j[t_{ij} \gamma_{ik}]-[\pd_j t_{ij}]\gamma_{ik} 
=\pd_j[t_{ij} \gamma_{ik}]+ f_{i}\gamma_{ik} \\
\label{W-v2}
m_{ij}[\pd_j \kappa_{ik}]&=\pd_j[m_{ij} \kappa_{ik}]-[\pd_j m_{ij}]\kappa_{ik} 
=\pd_j[m_{ij} \kappa_{ik}]-\epsilon_{ijl} t_{jl}\kappa_{ik}+ l_{i}\kappa_{ik} .
\end{align}
Now we add and subtract the term $\epsilon_{ikl}t_{ij}\kappa_{lj} $ in order to get the structure of the dislocation 
tensor~(\ref{Disl1}).
In addition, we introduce and use the following tensor
\begin{align}
\label{baregammma}
\bar\gamma_{ij}=\gamma_{ij}-\epsilon_{ijk}\phi_k=\pd_j u_i-\gamma_{ij}^{\text{P}}.
\end{align}
The purpose is to obtain an Eshelby stress tensor 
which is divergenceless in the limit to isotropic, source-free and homogeneous micropolar
elasticity.
We obtain the expression
\begin{align}
\label{J}
&\int\Big\{\epsilon_{kjl} t_{ij}\alpha_{il}
+\epsilon_{kjl} m_{ij}\Theta_{il}
-\epsilon_{kjl} t_{ji}\kappa_{li}^{\text{P}}
+f_i\bar\gamma_{ik}+l_i\kappa_{ik}
+\frac{1}{2}\, \gamma_{ij}[\pd_k A_{ijmn}]\gamma_{mn}
+\gamma_{ij}[\pd_k B_{ijmn}]\kappa_{mn}
\nonumber\\
&\hspace{6cm}
+\frac{1}{2}\, \kappa_{ij}[\pd_k C_{ijmn}] \kappa_{mn}
\Big\}\, \dd V \nonumber\\
&\quad
=\int\pd_j\big[w\delta_{jk}-t_{ij} \bar\gamma_{ik}
-m_{ij}\kappa_{ik}\big] \dd V=J_k .
\end{align}
The first integral contains the terms breaking the translational symmetry
called configurational forces.
The integrand of the second integral in Eq.~(\ref{J}) is the divergence 
of the (canonical) energy-momentum tensor (EMT) 
\begin{align}
\label{EMT1}
P_{kj}&=w\delta_{jk}-t_{ij}\bar\gamma_{ik}-m_{ij}\kappa_{ik},
\end{align}
which we call the Eshelby stress tensor of micropolar elasticity.
It is a generalization of the Eshelby stress tensor~\citep{Eshelby51,Eshelby75}
in elasticity towards micropolar elasticity.
Eq.~(\ref{EMT1}) is in agreement with the
so-called Maxwell stress tensor of Cosserat theory given by~\citet{Kluge,Kluge2}.
The divergence of the EMT~(\ref{EMT1}) can be integrated out 
with Gauss, and we find the $J$-integral for micropolar elasticity 
\begin{align}
J_k=\int  P_{kj}\, n_j \dd S .
\end{align}
The usual argument
of forming a penny shaped volume across the surface between two media 
I and II gives
\begin{align}
J_k=\int [P^{\text{I}}_{kj}-P^{\text{II}}_{kj}]\, n_j \dd S .
\end{align}
Here $P^{\text{I}}_{kj}$ and $P^{\text{II}}_{kj}$ 
are the energy-momentum tensors on the sides I and II of the interface, respectively.
The first integral in Eq.~(\ref{J})
defines a sum of configurational force densities:
\begin{align}
\label{F}
F_k=
\epsilon_{kjl}t_{ij}\alpha_{il}
+\epsilon_{kjl}m_{ij}\Theta_{il}
-\epsilon_{kjl} t_{ji}\kappa_{li}^{\text{P}}
+f_i\bar\gamma_{ik}+l_i\kappa_{ik}
+f^{\text{inh}}_k ,
\end{align}
where the inhomogeneity force density or Eshelby force density 
is due to the gradient of the
elastic tensors (see also~\citet{Eshelby51,Maugin93}):
\begin{align}
\label{f-inh}
f^{\text{inh}}_k=
\frac{1}{2}\, \gamma_{ij}[\pd_k A_{ijmn}]\gamma_{mn}
+\gamma_{ij}[\pd_k B_{ijmn}]\kappa_{mn}
+\frac{1}{2}\, \kappa_{ij}[\pd_k C_{ijmn}] \kappa_{mn}.
\end{align}
The first term in Eq.~(\ref{F}) is the configurational force density on a dislocation
density $\alpha_{il}$ in presence of a force stress $t_{ij}$,
\begin{align}
\label{PK1}
\epsilon_{kjl}t_{ij}\alpha_{il}.
\end{align}
This expression is the same as 
the Peach-Koehler force density in elasticity~\citep{PK}.
The second term is the configurational force density 
on a disclination density $\Theta_{il}$ in presence of a couple stress 
$m_{ij}$,
\begin{align}
\label{PK2}
\epsilon_{kjl}m_{ij}\Theta_{il}.
\end{align}
The expression~(\ref{PK2}) is the Mathisson-Papapetrou force density 
due to disclinations.
The third terms has a similar form as the Peach-Koehler force.
It is a configurational force density 
on a plastic wryness $\kappa^{\text{P}}_{li}$, caused by disclinations,
in presence of the force stress $t_{ji}$,
\begin{align}
\label{PK3}
-\epsilon_{kjl}t_{ji}\kappa_{li}^{\text{P}}.
\end{align}
These three configurational forces caused by defects have been already discovered
by~\citet{Kluge,Kluge2}.
The fourth term is the configurational force density
on a body force $f_i$ in presence of the distortion $\bar\gamma_{ik}$
\begin{align}
\label{Chere1}
f_i\bar\gamma_{ik} .
\end{align}
It is similar in form as the Cherepanov force density
(see, e.g., \citet{Cherepanov81,Eischen}).
The fifth term is the configurational force density
on a body couple $l_i$ in presence of an elastic wryness $\kappa_{ik}$ 
\begin{align}
\label{Chere2}
l_i\kappa_{ik} .
\end{align}

For homogeneous, source-free and compatible micropolar elasticity 
we recover the divergenceless Eshelby stress tensor as
\begin{align}
\label{EMT2}
P_{kj}&=w\delta_{jk}-t_{ij}\pd_i u_k-m_{ij}\pd_i \phi_{k},
\qquad
\pd_j P_{kj}=0.
\end{align}
This formula is in agreement with the Eshelby stress tensor
given by~\citet{LM03}.
The corresponding Eshelby stress tensor for finite theory of polar elasticity
has been given by~\citet{Maugin98} and \citet{NZ}.

\section{The angular momentum tensor (AMT) and configurational moments in micropolar
  elasticity}
\setcounter{equation}{0}

Now, having the AMT in mind, we specify the functional derivative 
to be rotational:
\begin{align}
\label{rot}
\delta=(\delta x_k)\epsilon_{kji} x_j\pd_i ,
\end{align} 
where $(\delta x_k)$ denotes the $x_k$-direction of the axis of rotation.
Using the same manipulations as in section~\ref{EMT-sec}, we find:
\begin{align}
\label{L0}
\epsilon_{kji}x_j J_i =
\int\epsilon_{kji}x_j F_i\, \dd V
=\int\epsilon_{kji}\big[\pd_n(x_j P_{in})-P_{ij}\big] \dd V
\end{align}
with Eq.~(\ref{F}).
Now we rewrite the part 
\begin{align}
\epsilon_{kji} P_{ij}=-\epsilon_{kji}(t_{lj}\bar\gamma_{li}+m_{lj}\kappa_{li}).
\end{align}
When we 
subtract and add the terms $\epsilon_{kji}t_{il}\bar\gamma_{jl}$
and $\epsilon_{kji} m_{li}\kappa_{lj}$, and use the 
equilibrium equations~(\ref{fe1}) and (\ref{fe2}), and the decompositions
of the strain tensors~(\ref{deco1}) and (\ref{deco2}), 
we obtain the result
\begin{align}
\label{L1}
&\int\epsilon_{kji}\Big\{
x_j F_i
+f_j u_i+l_j\phi_i
+t_{il}\gamma_{jl}^{\text{P}}+m_{il}\kappa_{jl}^{\text{P}}
+t_{il}\bar\gamma_{jl}-t_{lj}\bar\gamma_{li}
+m_{il}\kappa_{jl}-m_{lj}\kappa_{li}
+\epsilon_{ilk}t_{lk}\phi_j\Big\}\dd V \nonumber\\
&\quad
=\int\epsilon_{kji}\pd_n \big[x_j P_{in}+u_j t_{in}+\phi_i m_{in}\big] \dd V= L_k .
\end{align}
The first integral contains terms breaking the rotational symmetry
which we call configurational moment densities.
The first term is a configurational vector moment 
produced by the configurational
forces $F_j$. 
The other terms are intrinsic vector moments
as a result of the manipulation during the calculation
in order to obtain an angular momentum tensor.
The integrand of the second integral in Eq.~(\ref{L1}) is the
divergence of the (canonical) AMT
\begin{align}
\label{AMT2}
M_{kn}=\epsilon_{kji}\big[x_j P_{in}+u_j t_{in}+\phi_j m_{in}\big].
\end{align}
This is the incompatible generalization of the compatible result given by~\citet{PS,LM03}.
Eq.~(\ref{AMT2}) consists of two parts:
the first one is the orbital AMT given in terms of the EMT and the
second one is the spin AMT given in terms of the force and couple stresses. 
Eq.~(\ref{L1}) may be transformed into a surface integral
\begin{align}
L_k =\int M_{kj}\, n_j \dd S .
\end{align}

The tensor $M_{ki}$ is related to a tensor of third rank as follows:
\begin{align}
M_{kn}=\frac{1}{2}\, \epsilon_{kji} M_{jin}.
\end{align}
It may be decomposed according
\begin{align}
M_{jin}=L_{jin}+S_{jin} ,
\end{align}
with the orbital angular momentum tensor
\begin{align}
L_{jin}=x_j P_{in}-x_i P_{jk}
\end{align}
and the spin angular momentum tensor
\begin{align}
S_{jin}=u_j t_{in}-u_i t_{jn}
+\phi_{j} m_{in}-\phi_{i}m_{jn}.
\end{align}
It is the canonical (or Noether) spin AMT.
Furthermore, the spin part can be written as:
\begin{align}
S_{jin}=u_\alpha (\Sigma_{ji})^{\alpha\beta}\, t_{\beta n}
+\phi_{\alpha } (\Sigma_{ji})^{\alpha\beta}\, m_{\beta n},
\end{align}
with the infinitesimal generator of the finite-dimensional irreducible
representation 
of the rotation group for a vector field (spin-1 field):
\begin{align}
(\Sigma_{ji})^{\alpha\beta}=\delta_j^\alpha\delta_i^\beta-\delta_i^\alpha\delta_j^\beta .
\end{align}

Since
\begin{align}
\pd_j M_{kj}\neq 0,
\end{align}
the configurational moments break the rotational symmetry.
Only for source-less ($f_i=0$, $l_i=0$), isotropic, compatible and homogeneous
micropolar elasticity, the AMT is divergenceless.
The isotropy condition is:
\begin{align}
\label{IC2}
&\epsilon_{kji}\big[
t_{il}\bar\gamma_{jl}-t_{lj}\bar\gamma_{li}
+m_{il}\kappa_{jl}-m_{lj}\kappa_{li}+\epsilon_{ilk}t_{lk}\phi_j
\big]=0 .
\end{align}
Eq.~(\ref{IC2}) has to be fulfilled by the isotropic 
constitutive relations~(\ref{CR1}), (\ref{CR2}) and (\ref{CR-Iso}). 
It is the generalization of the isotropy condition of elasticity (see, e.g.,~\citet{Eshelby75}).

\section{The scaling flux and configurational work in micropolar elasticity}
\setcounter{equation}{0}
Having the $M$-integral in mind, we specify the functional derivative 
to be dilatational:
\begin{align}
\label{dil}
\delta=  x_k \pd_k .
\end{align} 
Using Eq.~(\ref{J}), we find
\begin{align}
\label{M0}
x_k J_k=\int x_k F_k\, \dd V
=\int\big[\pd_j(x_k P_{kj})-P_{kk}\big] \dd V
\end{align}
with
\begin{align}
P_{kk}=\pd_j\Big[\frac{n-2}{2}\, u_i t_{ij}+\frac{n}{2}\, \phi_i m_{ij}\Big]
+\frac{n-2}{2}\, (f_i u_i-t_{ij}\gamma_{ij}^{\text{P}})
+\frac{n}{2}\, (l_i \phi_i-m_{ij}\kappa_{ij}^{\text{P}})
-m_{ij}\kappa_{ij}
\end{align}
and $\delta_{kk}=n$. Thus, $n=3$ for three dimensions and $n=2$ for two
dimensions.
The $M$-integral generalized for micropolar elasticity 
of an anisotropic, non-homogeneous medium with body forces and couples is
\begin{align}
\label{M1}
&\int \Big\{ x_k F_k
+\frac{n-2}{2}\, f_i u_i+\frac{n}{2}\, l_i \phi_i
-\frac{n-2}{2}\, t_{ij}\gamma_{ij}^{\text{P}}-\frac{n}{2}\, m_{ij}\kappa_{ij}^{\text{P}}
-m_{ij}\kappa_{ij}
\Big\}\, \dd V \nonumber\\
&\quad
=\int\pd_j \Big[x_k P_{kj}-\frac{n-2}{2}
\,u_k t_{kj}-\frac{n}{2}\,\phi_{k}m_{kj} \Big] \dd V= M .
\end{align}
It can be seen that the (configurational work) terms 
appearing in the first integral in Eq.~(\ref{M1}) 
break the dilatation (or scaling) invariance. 
The first term is built from the configurational forces by multiplication
with $x_k$. The other terms are called intrinsic scalar moments.
The integrand of the second integral in Eq.~(\ref{M1}) is the
divergence of the scaling flux vector
\begin{align}
\label{dil-v}
Y_{j}=\Big[x_k P_{kj}-\frac{n-2}{2}\, u_k t_{kj}-\frac{n}{2}\, \phi_{k} m_{kj}\Big].
\end{align}
Eq.~(\ref{dil-v}) is the incompatible generalization of the compatible 
result given by~\citet{LM03}. 
The term $-(n-2)/2$ is the scale (or canonical) dimension of the displacement vector $u_k$
and  $-n/2$ is the scale (or canonical) dimension of the axial vector field $\phi_k$.
The first term in Eq.~(\ref{dil-v}) is the `orbital' piece and the 
other two terms are the `intrinsic' parts of the scaling flux vector.

Eq.~(\ref{dil-v}) can be transformed into a surface integral
\begin{align}
M =\int Y_{j}\, n_j \dd S .
\end{align}
In general, the dilatation current is not divergenceless
\begin{align}
\pd_j Y_j\neq 0 .
\end{align}
Therefore, in this case the scaling symmetry is broken.
Even in the compatible, homogeneous and source-free case the 
scaling symmetry is broken; namely
\begin{align}
M=-\int m_{ij}\kappa_{ij}\, \dd V.
\end{align}
The reason is that field theories like 
gradient elasticity, micropolar elasticity and micromorphic elasticity
are theories with internal length scales.
Because the material tensors have different dimensions, 
such constants with the dimension of length appearing in the 
Lagrangian (strain energy density)
violate the dilatational (scaling) invariance.

\section{Conclusion}
In this paper, we have derived broken conservation laws of micropolar elasticity
by using the framework of the Noether theorem on invariant variational
principles. 
Earlier results obtained by~\citet{Kluge,Kluge2,PS,LM03} have been extended to 
account for material nonhomogeneity, anisotropy, defects (dislocations, disclinations),
body forces and body couples. 
We calculated the Eshelby stress tensor, angular momentum tensor and scaling flux vector,
which are not divergenceless, for such an extended micropolar theory. 
Additionally, we have given the $J$-, $L$- and $M$-integrals 
for this extension.
The terms breaking the translational, rotational and scaling invariance
are called configurational forces, moments and work, respectively.

\subsection*{Acknowledgement}
M.L. has been supported by an Emmy-Noether grant of the 
Deutsche Forschungsgemeinschaft (Grant No. La1974/1-2).

\end{document}